# Novel Bismaleimide Resin/Silsesquioxane and Titania Nanocomposites by the Sol-Gel Process: the Preparation, Morphology, Thermal and Thermo-mechanical Properties

Guotao Lu[a, b,*], Ying Huang[a]

[a]*Center for Molecular Sciences, Institute of Chemistry, Chinese Academy of Sciences, Beijing 100080, People's Republic of China*
[b]*Research and Development Center, Bonna-Agela Technologies, Inc. Wilmington, DE 19808*

---

[*] Corresponding author. Tel: +1-302-690-7160

*E-mail address*: luguotao@gmail.com (G. Lu).

**Abstract**

Bismaleimide(BMI) resin/silsesquioxane or titania nanocomposites were synthesized from bismaleimide resin and $SiO_{3/2}$ or $TiO_2$ via the sol-gel process of N-γ-triethoxylsilylpropyl-maleamic acid (TESPMA) or tetrabutyltitanate ($Ti(O^nBu)_4$, TBT), respectively, in the presence of the AP-BMI prepolymers. These nanocomposite materials were characterized by FT-IR, FE-SEM, TGA and DMA. It was found that the nano-scale $SiO_{3/2}$ or $TiO_2$ particles were formed in the AP-BMI resin matrix and the average original particle size of the dispersed phase in the nanocomposites was less than 100nm, but the particle aggregates with bigger size existed. Obvious improvements of $T_g$ and the heat resistance of the AP-BMI resin were achieved by introduction of the nano-sized $SiO_{3/2}$ inorganic phase, and the modulus at high temperatures was improved too. The incorporation of nano-scale $TiO_2$ particles into the AP-BMI resin improved the $T_g$ of the polymer, but lowered the thermal resistance of the material, and improved the modulus of the material at lower temperatures, but lowered the modulus at higher temperatures.

# 1. Introduction

There has been increasing interest in Organic-Inorganic Nanocomposite (OINC) materials field due to their wide and potential applications in electronics, optics, chemistry, and biomedicine [1-4]. The incorporation of inorganic component into organic polymer matrix has greatly improved the thermal and mechanical properties of the polymer. And careful design of the inorganic phase can introduce many new properties into the polymers, such as catalysis [5], photo refraction [6], electroluminescence [7], and photochromophore [8], etc. The commonly used inorganic phases are nano-scale metal particles [9], metal oxides [10], semiconductor nanocrystals [11], carbon nanotube [12], and mesoporous organosilica [1].

The sol-gel process, which produces glasses and ceramics at a relatively low temperature rather than at very high temperature needed for conventional inorganic glasses and ceramics, is a convenient method to be used to prepare OINC [10]. Soluble polymers or their monomers are usually introduced into the sol-gel system to provide the organic component in OINC [10]. Organic-inorganic nanocomposite materials of polymethacrylates [13], polyacrylonitrile [14], polystyrene [15], polycarbonate [16], epoxy resin [17], phenolic resin [18] and polymaleimide [19] with $SiO_2$ or $TiO_2$ nano-particles have been prepared through the sol-gel process.

The resultant nanocomposites can be divided into two classes according to the interfacial nature: one is nanocomposites with interfacial covalent bonds or ionic-covalent bonds between the organic and inorganic phases [13, 15]; the other is nanocomposites without covalent bonds between the two phases, only with hydrogen bonds or Van Der Walls interaction between them [20-22]. Generally, the nanocomposite material with covalent bonds between the two phases has better thermal and mechanical properties than those without covalent bonds [23].

High performance thermosetting resins have wide applications in many areas such as aerospace and semiconductor industries. To improve their properties further, nanocomposite materials based on high performance thermosetting resins have been studied, such as PMR-15 with clay [24], polybenzoxazine with clay [25] and $TiO_2$ [26], Novolac resin with $SiO_2$ [27] and $TiO_2$ [28], and cyanate ester with clay[29], etc.

Bismaleimide resin is one of the most important classes of high performance thermosetting resins, with its outstanding properties including high tensile strength and modulus, excellent heat, chemical and corrosion resistance [30, 31]. However, to overcome its brittleness (the major disadvantage) and improve its thermomechamical properties, the performance enhancement is needed. Recently, various research has been done to overcome BMI resins' shortcoming and enhance its thermal and mechanical properties, for example, the studies of BMI-containing novel monomers [32-33], its copolymers [34-35], polymer alloys [36-39], carbon fiber [40-41] and Al-B whiskers [42] reinforced composites.

Performance enhancement of BMI resin has also been studied by preparing its nanocomposite material with clay [43] and $TiO_2$ (polyetherimide as BMI's modifier) [44]. In the present work we reported the preparation of BMI resin/$SiO_{3/2}$ or $TiO_2$ nanocomposites by the sol-gel process and studied the morphology, thermal and mechanical properties of the resultant materials, especially compared the effect of the different interfaces (different types of covalent bonds) to the properties of the nanocomposite materials. To improve the toughness and reduce the viscosity of BMI melt, allyl novolac phenolic resin (AP) has been used as BMI's modifier.

## 2. Experimental

*2.1. Materials.*

Phenol (CP) and formaldehyde (37% in water, CP) were from Beijing Organic Chemicals Factory (China), and were used as supplied. Allyl chloride was obtained from Qilu Petrochemical Co. (China) and used after distillation. 4,4′-Bismaleimidodiphenyl methane (BMI) was purchased from Fenguang Chemical Co., Ltd. (China). It is a crystalline substance with melting point of 151-154°C and the purity greater than 99%. Tetrabutyl titanate (Ti(O$^n$Bu)$_4$, TBT) was a chemical pure reagent from Tianjian 2nd Chemicals Factory (China), and was used as supplied. Acetyl acetone (ACAC), Tetrahydrofuran (THF) and n-Butanol (BuOH) were analytical reagents from the Beijing Chemical Factory (China). THF was purified by reflux over Na/CO(C$_6$H$_5$)$_2$ and distillation. De-ionized water was acidified by hydrochloric acid (35-38%) to pH=2.

N-γ-Triethoxylsilylpropyl-maleamic acid (TESPMA) was prepared according to a method described in our previous work [19]. The allylated novolak resin (AP) was prepared according to the method described in the reference [45]. Molecular weight of AP-1 and AP-2 was 1180 and 1075, respectively, and the degree of allylation was 39% and 59% determined by $^1$H-NMR.

*2.2. Synthesis of The ally-modified novolak resin-BMI prepolymers(AP-BMI).*

The allylated Phenol resins (AP) were prepared according to the literature [45]. For AP-BMI-1-TSi-X, 58.3g of AP (containing 0.17mol allyl groups) was charged in a flask and heated to 120°C. 13.2g TESPMA (containing 0.03mol unsaturated groups) was added under stirring. The reaction mixture was kept at 120°C for 9hrs. Then, 24.8g BMI (0.14mol) was introduced. The reaction was continued for 80mins to give a transparent red-brown BMI-modified novolac resin bearing triethoxysilylpropyl groups (AP-BMI-1-TSi-2). The formation of AP-BMI-1-TSi-2 was confirmed by FT-IR and $^1$H-NMR (CDCl$_3$) spectra. The compositions of AP-BMI-1-TSi-2 prepolymers are listed in Table 1. $^1$H-NMR (CDCl$_3$, 300M, δ ppm):

Anal. Calcd for AP-BMI-1-TSi-2 (

To synthesize AP-BMI-2, 61.0g of APN (containing 0.25mol allyl groups) was charged in a three-necked flask and heated to 120°C. Then, 44.8g BMI (0.25mol) was introduced. The reaction was continued for 2hrs to give a transparent red-brown AP-BMI-2 prepolymer. The formation of AP-BMI-2 was also confirmed by FT-IR and $^1$H-NMR (CDCl$_3$) spectra.

*2.3. Synthesis of the AP-BMI-1/SiO$_{3/2}$ nanocomposites.*

AP-BMI-1/SiO$_{3/2}$ nanocomposites were synthesized through sol-gel reactions of AP-BMI-1-TSi catalyzed by hydrochloric acid, and the subsequent curing initiated by heat. The following was a typical procedure for preparation of the nanocomposite AP-BMI-1/Sil-2: 188.7g AP-BMI-1-TSi-2 was charged in a three-neck round flask equipped with a mechanical stirrer and a reflux condenser. Then 46g THF was added, and the system was heated to 70°C with an oil bath. After the prepolymer was dissolved completely, 1.7ml of hydrochloric acid (0.01M) was added under intensive stirring to carry out the sol-gel reactions. The reaction was continued at 70°C for 4hrs. Then the volatile was removed at 50-130°C under reduced pressure to give the mold powder of the nanocomposites.

Specimens of the nanocomposites were prepared by compression molding. The curing cycle was 140°C/20MPa/4h + 200°C/20MPa/6h. The post cure schedule was 250°C/6h. The compositions of the AP-BMI-1/SiO$_{3/2}$ nanocomposites are listed in Table 2.

*2.4. Synthesis of the AP-BMI-2/TiO$_2$ nanocomposites.*

AP-BMI-2/TiO$_2$ nanocomposite materials were prepared by using the following procedure: 8.16g BMI-PN (Allylation degree is 59.2%) was weighed into a three-necked round flask equipped with a mechanical stirrer and a reflux condenser. Then 6.23g THF was added into the flask and the system was heated in an oil bath to keep the mixtures' temperature at 80°C. After the prepolymer was dissolved completely, the clear mixture of 0.68g Ti(O$^n$Bu)$_4$ and 0.3g acetyl acetone was added to the solution, and at last 0.03g acidified water with pH=2.0 was added to the mixture to perform the sol-gel reaction. The reaction time was 1.5hrs and the system's temperature was kept at 80°C. Then the volatile was removed at 50-130°C under reduced pressure to give the mold powder of the nanocomposite.

Specimens of the nanocomposites were prepared by compression molding. The curing cycle was 140°C/20MPa/4h + 200°C/20MPa/6h. The post cure schedule was 250°C/6h. The compositions of the AP-BMI-2/TiO$_2$ nanocomposites are listed in Table 2.

*2.5. Measurements.*

$^1$H NMR spectra were obtained with a Bruker MW 300 spectrometer with CDCl$_3$ as solvent. FT-IR spectra were recorded on a Perkin-Elmer model 1600 IR spectrometer. Dynamical mechanical analysis (DMA) was performed on a Perkin-Elmer DMA-7 Dynamical Mechanical Analyser in the bending mode with the specimen dimension 15mm×4mm×1.5mm. The measurements were conducted at 2Hz in the temperature range from 40°C to 350°C at heating rate 20°C/min. Thermogravimetric analysis (TGA) was performed on a Perkin-Elmer TGA-7 thermogravimeter with a heating rate of 20°C/min under a nitrogen flow rate of 100 ml/min. The microstructure features of the nanocomposite materials were examined with a Hitachi S-900 field emission scanning electron microscope. The fracture surfaces of the samples were coated with gold to eliminate charging effects, and a high voltage (10KV) was used.

## 3. Results and discussion

*3.1. Synthesis of the AP-BMI/SiO$_{3/2}$ and TiO$_2$ nanocomposites.*

*3.1.1 Synthesis of the AP-BMI-1/SiO$_{3/2}$ nanocomposites.*

Scheme 1 gives the synthesis schedule of the AP-BMI-1/SiO$_{3/2}$ nanocomposites. It can be seen that for the synthesis of the AP-BMI-1/SiO$_{3/2}$ nanocomposites, N-γ-triethoxylsilylpropyl-maleamic acid (TESPMA) was introduced into AP-BMI-1 prepolymer chain by its "Ene" reaction with allyl groups of allylated phenolic (AP) modified BMI, and through its hydrolysis and condensation, SiO$_{3/2}$ inorganic phase was incorporated into BMI resin matrix.

FT-IR spectroscopy has been used to monitor the above process. As shown by the spectra in Figure 1 and 2 for the AP-BMI-1-TSi prepolymer (a) and the corresponding nanocomposite AP-BMI-1/SiO$_{3/2}$ (b), we can see that the adsorption at 827cm$^{-1}$ (Fig. 2c), 1076cm$^{-1}$(Fig. 2c), 1175cm$^{-1}$(Fig. 2c), 2965 and 2875cm$^{-1}$ (Fig. 2a) from the Si-OC$_2$H$_5$ almost completely disappeared after the sol-gel reaction and the curing process. An absorbance band associated with Si-OH centered at 917cm$^{-1}$(Fig. 2c) and the band at 1100cm$^{-1}$(Fig. 2c) and 464cm$^{-1}$ (Fig. 2d) from SiO$_{3/2}$ networks appeared. All above changes substantiated that the ethoxy groups bound to Si atoms were hydrolyzed to Si-OH groups and then Si-O-Si networks were formed by the condensation of the latter. The formation of SiO$_{3/2}$ networks can also be proved by SEM result.

*∗Figure 1 should be inserted here.*

*∗Figure 2 should be inserted here.*

The ring-closure reaction of N-γ-triethoxylsilylpropyl-succinamic acid bonded to the AP-BMI-1 resin matrix (Scheme 1, formula (A)) occurred as indicated by disappearance of the absorption at 1560, 1544 and 1297cm$^{-1}$(Fig. 2b) from amide groups during the curing process, the shift of carbonyl group's adsorption from 1711 to 1706cm$^{-1}$(Fig. 2b), and the increase of the adsorption from O=C-N-C=O group(centered at 1780cm$^{-1}$(Fig. 2b)). This ring-closure step happened in the curing process is helpful to enhance the thermal resistance of the nanocomposites.

The proposed curing mechanism of the AP-BMI-1/SiO$_{3/2}$ nanocomposites is similar to DABA-BMI system given by Sung and Phelan [46] (Figure 1 and 2 there), and it includes Ene, Diels-Alder, homopolymerization, rearomatization, and alternating copolymerization [46]. Those reactions can be proceeded by the disappearance of the absorptions at 3100cm$^{-1}$(Fig. 2a), the significant reduction of the bands from allyl groups at 990 and 930cm$^{-1}$(Fig. 2c) and the disappearance of the absorption at 950cm$^{-1}$ (Fig.2c)from the double bonds in BMI ring, which also resulted in the shift of the carbonyl band from 1711 to 1706cm$^{-1}$ (Fig. 2c). A new adsorption band associated with C-N-C succinimide centered at 1180cm$^{-1}$ (Fig. 2c) appears while the adsorption band at 1150cm$^{-1}$ (Fig. 2c) associated with C-N-C maleimide decreases remarkably. Based on above results and the fact that the BMI resin was cured at 250°C, we believe that the AP-BMI-1 resin could be cured completely, according to the result by Phelan and Sung [46]. This accounts for the good thermal and mechanical properties of the composite resin.

*3.1.2 Synthesis of the AP-BMI-2/TiO$_2$ nanocomposites.*

The synthesis process of the AP-BMI-2/TiO$_2$ nanocomposites has been shown in Scheme 2. To reduce the hydrolysis rate of tetrabutyltitanate (TBT), AcAc has been used to complex with Ti atoms. The hydrolysis of TBT gives Ti-OH groups, which condensed into TiO$_2$ nano-particles at BMI resin matrix. From Scheme 2, we can see that most TiO$_2$ nano-particles are not covalent bonded into BMI resin network, whereas as shown in Scheme 1, $SiO_{3/2}$ network are covalent bonded to BMI resin network. This difference determines their big contrast in thermal and mechanical properties.

Figure 3 and 4 give the FT-IR spectra for the AP-BMI-2 prepolymer (a) and the corresponding nanocomposite AP-BMI-2/$TiO_2$ (b). The formation of $TiO_2$ particles can be proofed by the appearance of a new adsorbance band at 670 cm$^{-1}$ (Fig. 4c) from Ti-O-Ti in $TiO_2$, whereas the presence of Ti-O-C peak at 827 cm$^{-1}$ (Fig. 4c) shows that the hydrolysis of TBT is not complete. The formation of $TiO_2$ nanoparticles can also be seen from SEM study.

*Figure 3 should be inserted here.*

*Figure 4 should be inserted here.*

The proposed curing mechanism of the nanocomposite AP-BMI-2/$TiO_2$ is the same with the AP-BMI-1/$SiO_{3/2}$ nanocomposites. The main changes of FT-IR spectra have been shown in Figure 4, which are very similar to those of the FT-IR spectra of the AP-BMI-1/$SiO_{3/2}$ nanocomposites (Figure 2), except for no ring-closure reaction here.

*3.2. Morphology of the AP-BMI/$SiO_{3/2}$ and $TiO_2$ nanocomposites.*

The morphology of the AP-BMI/$SiO_{3/2}$ and $TiO_2$ nanocomposites has been studied by FE-SEM. Figure 5 gives the FE-SEM photographs of the fracture surface of the AP-BMI/$SiO_{3/2}$ and $TiO_2$ nanocomposites. It can be seen that the AP-BMI resin matrix was a homogeneous material and no phase separation was observed (Figure 5, a and d), but the AP-BMI/$SiO_{3/2}$ and $TiO_2$ nanocomposites were two-phase materials, in which the continuous phases were AP-BMI resin, and the dispersed phases were consisted of the silsesquioxane network (Figure 5, b and c) and $TiO_2$ particles (Figure 5, e and f), respectively. The average diameters of original particles of the dispersed phase in the nanocomposites were ca. 100nm, with particle aggregates of bigger sizes when the inorganic content increasing (Figure 5, b and c for $SiO_{3/2}$, e and f for $TiO_2$).

*Figure 5 should be inserted here.*

*3.3. The thermal and mechanical properties of the the AP-BMI/SiO$_{3/2}$ and TiO$_2$ nanocomposites.*

*3.3.1 $T_g$'s of the AP-BMI/SiO$_{3/2}$ and TiO$_2$ nanocomposites.*

Glass transition temperatures ($T_g$'s) of the AP-BMI/SiO$_{3/2}$ and TiO$_2$ nanocomposites are characterized by DMA, shown in Figure 6. Figure 7 gives the $T_g$'s of the AP-BMI/SiO$_{3/2}$ and TiO$_2$ nanocomposites changes with inorganic content based on the data from Figure 6. It can be seen in Figure 7 that the incorporation of SiO$_{3/2}$ nanoparticles improved the glass transition temperature of the AP-BMI resin due to the interaction of SiO$_{3/2}$ nanoparticles and AP-BMI molecular chain, where the SiO$_{3/2}$ behave as physical cross-link points, or forming SiO$_{3/2}$/AP-BMI interpenetrate networks. Both factors will limit the molecular chain movement of AP-BMI resin, thus increase its glass transition temperature. For the AP-BMI/TiO$_2$ nanocomposites, the introduction of TiO$_2$ nano-particles into AP-BMI resin had obvious effect in improving the $T_g$'s of AP-BMI resin. Here there is no covalent bonds between the resin and the inorganic phase, then the TiO$_2$ behaving as physical cross-link points is the main reason for the improvement of the resin's glass transition temperature. At Zhao et al.' work [44], the introduction of TiO$_2$ also improved PEI-BMI resin's $T_g$, consisted with this paper, so it seems that the introduction of TiO$_2$ has eminent effect to increase $T_g$ of the resin in both works.

*Figure 6 should be inserted here.*

*Figure 7 should be inserted here.*

*3.3.2 Thermal resistance of the AP-BMI/SiO$_{3/2}$ and TiO$_2$ nanocomposites.*

TGA diagrams of the AP-BMI/SiO$_{3/2}$ and TiO$_2$ nanocomposites with different inorganic phase contents are shown in Figure 8. It is seen from the figure that there were two stages in the weight-loss process of the nanocomposites. The elimination of water formed by the condensation of residual Si-OH or Ti-OH groups (most Si-OH and Ti-OH groups have been condensed at post-cure process) or uptake moisture accounted for the minor weight loss started at ~150$^{o}$C. The major weight loss occurred at >450$^{o}$C, which was due to the decomposition of the AP-BMI resin in the AP-BMI/SiO$_{3/2}$ and TiO$_2$ nanocomposites. Table 3 summarizes the temperatures for 5% weight loss, 10% weight loss, and the temperatures at maximum decomposition rate (T$_{max}$) of the AP-BMI/SiO$_{3/2}$ and TiO$_2$ nanocomposites in the TGA diagrams.

*Figure 8 should be inserted here.*

It is seen that the introduction of the SiO$_{3/2}$ inorganic phase into the AP-BMI resin matrix improved the temperatures for 5% weight loss, 10% weight loss, the T$_{max}$ and the weight retention at 700 $^{o}$C of the resin. The reasons accounted for the increase of the glass transition temperature of the AP-BMI-1/SiO$_{3/2}$ nanocomposites also hold true for this situation. However, the introduction of the TiO$_2$ into the AP-BMI resin lowered the temperatures for 5% weight loss, 10% weight loss, the T$_{max}$ and the weight retention at 700 $^{o}$C of the resin. The probable reasons lie in that except there were no covalent bonds between organic phases and inorganic phases, the other is the remained AcAc could catalyze the decomposition process of the AP-BMI resin matrix, which acts as a solvent of the resin.

*3.3.3 Dynamical moduli of the AP-BMI/SiO$_{3/2}$ and TiO$_2$ nanocomposites.*

The change in storage modulus with temperature for the AP-BMI/SiO$_{3/2}$ and TiO$_2$ nanocomposites has been shown in Figure 9. It can be seen from the figure that the moduli of the nanocomposites at high temperatures improved significantly due to the introduction of the SiO$_{3/2}$ nano-phase into the resin matrix. The main reason here is that the covalent bonds between the organic phase and inorganic phase in the nanocomposites limit the molecular movement of the AP-BMI resin. While the introduction of the titania phase into the AP-BMI resin improved the moduli of the nanocomposites at lower temperatures (<200°C), but lowered the moduli at higher temperatures (>250°C). The explanation above for the decrease in the thermal resistance also accounts for this phenomenon.

∗*Figure 9 should be inserted here.*

## 4. Conclusions

The AP-BMI/SiO$_{3/2}$ and TiO$_2$ nanocomposites have been prepared through the hydrolysis and condensation reactions of –Si(OEt)$_3$ and TBT in the presence of AP-BMI prepolymers, respectively. The nanometer SiO$_{3/2}$ or TiO$_2$ particles were formed in the BMI resin matrix and the average original particle size of the dispersed phase in the nanocomposites was less than 100nm, but the particle aggregates with bigger size existed. Because the covalent bonds connected the organic and inorganic phases existed in the AP-BMI-1/SiO$_{3/2}$ nanocomposites, and did not exist in the AP-BMI-2/TiO$_2$ nanocomposites, the thermal and mechanical properties of the two nanocomposites are very different. The introduction of the nanosized SiO$_{3/2}$ inorganic phase into AP-BMI resin improved $T_g$, the heat resistance of the BMI resin, and its modulus at high temperatures remarkablely. However, the incorporation of nano-scale TiO$_2$ particles into the AP-BMI resin improved the $T_g$ of the resin, but lowered the thermal resistance and the modulus at higher temperatures of the material, although the modulus of the material at lower temperatures has been improved.

Table 1

Compositions of the APN-BMI-TSi prepolymers.

| Code | AP (wt-%) | TSi (wt-%) | BMI (wt-%) | Solid content of BMI-TSi-PN (wt-%) |
|---|---|---|---|---|
| AP-BMI-1 | 63.2 | 0 | 36.8 | 92.9 |
| AP-BMI-1-TSi-2 | 56.4 | 11.0 | 32.6 | 90.9 |
| AP-BMI-1-TSi-4 | 48.5 | 23.4 | 28.1 | 90.4 |

Table 2

Compositions of the AP-BMI resin/$SiO_{3/2}$ or $TiO_2$ nanocomposites.

| Code | $SiO_2/TiO_2$ content (wt-%) | | AP (wt-%) | BMI (wt-%) | AP/BMI by wt-% |
|---|---|---|---|---|---|
| | $SiO_2$(wt-%) | R*(wt-%) | | | |
| AP-BMI-1 | 0 | 0 | 63.2 | 36.8 | 1.72:1 |
| AP-BMI-1/Sil-2 | 1.9 | 5.0 | 58.9 | 34.2 | 1.72:1 |
| AP-BMI-1/Sil-4 | 4 | 10.7 | 54.0 | 31.3 | 1.72:1 |
| AP-BMI-2 | | 0 | 54.6 | 45.4 | 1.20:1 |
| AP-BMI-2/Tit-2 | | 2 | 53.5 | 43.5 | 1.20:1 |
| AP-BMI-2/Tit-5 | | 5 | 51.9 | 43.1 | 1.20:1 |

**\* R stands for** 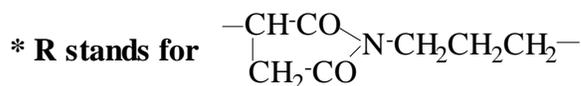

Table 3

Thermal resistantce of the AP-BMI resin/$SiO_{3/2}$ or $TiO_2$ nanocomposites.

| Sample | Inorganic phase content (wt-%) | Temp- for 5% wt- loss (°C) | Temp- for 10% wt- loss (°C) | $T_{max}$ (°C) | Wt- retention at 700°C (%) |
|---|---|---|---|---|---|
| AP-BMI-1 | 0 | 331 | 408 | 636 | 19.4 |
| AP-BMI-1/Sil-2 | 1.9 | 384 | 432 | 653 | 33.5 |
| AP-BMI-1/Sil-4 | 4 | 407 | 444 | 672 | 39.1 |
| AP-BMI-2 | 0 | 415 | 452 | 655 | 42.5 |
| AP-BMI-2/Tit-2 | 2 | 419 | 438 | 540 | 35.9 |
| AP-BMI-2/Tit-5 | 5 | 425 | 428 | 520 | 35.8 |

**Figure Caption**

Scheme 1.　The prepareation of the AP-BMI resin/SiO$_{3/2}$ nanocomposites.

Scheme 2.　The prepareation of the AP-BMI resin/TiO$_2$ nanocomposites.

Figure 1.　FT-IR spectra for for the AP-BMI resin/SiO$_{3/2}$ nanocomposites: (a) APN-BMI-1-TSi; and (b) APN-BMI/Sil-2.

Figure 2.　Magnification of Figure 1.

Figure 3.　FT-IR spectra for for the AP-BMI resin/SiO$_{3/2}$ nanocomposites: (a) APN-BMI-2; and (b) APN-BMI/Tit-2.

Figure 4.　Magnification of Figure 3.

Figure 5.　FE-SEM images for the AP-BMI resin/SiO$_{3/2}$ or TiO$_2$ nanocomposites (the scale bar is 2μm): (a) cured AP-BMI-1; (b) AP-BMI/Sil-2; (c) AP-BMI/Sil-4; (d) cured AP-BMI-2; (e) AP-BMI-2/Tit-2; and (f) AP-BMI-2/Tit-5.

Figure 6.　Glass transition temperatures of the AP-BMI resin/SiO$_{3/2}$ or TiO$_2$ nanocomposites measured by DMA.(a) cured AP-BMI-1; (b) AP-BMI/Sil-2; (c) AP-BMI/Sil-4; (d) cured AP-BMI-2; (e) AP-BMI-2/Tit-2; and (f) AP-BMI-2/Tit-5.

Figure 7.　Glass transition temperatures of the AP-BMI resin/SiO$_{3/2}$ or TiO$_2$ nanocomposites changes with inorganic component content.

Figure 8.　TGA diagrams of the AP-BMI resin/SiO$_{3/2}$ or TiO$_2$ nanocomposites: (a) cured AP-BMI-1; (b) AP-BMI/Sil-2; (c) AP-BMI/Sil-4; (d) cured AP-BMI-2; (e) AP-BMI-2/Tit-2;

and (f) AP-BMI-2/Tit-5.

Figure 9. Change of storage modulus with temperature for the AP-BMI resin/$SiO_{3/2}$ or $TiO_2$ nanocomposites: (a) cured AP-BMI-1; (b) AP-BMI/Sil-2; (c) AP-BMI/Sil-4; (d) cured AP-BMI-2; (e) AP-BMI-2/Tit-2; and (f) AP-BMI-2/Tit-5.

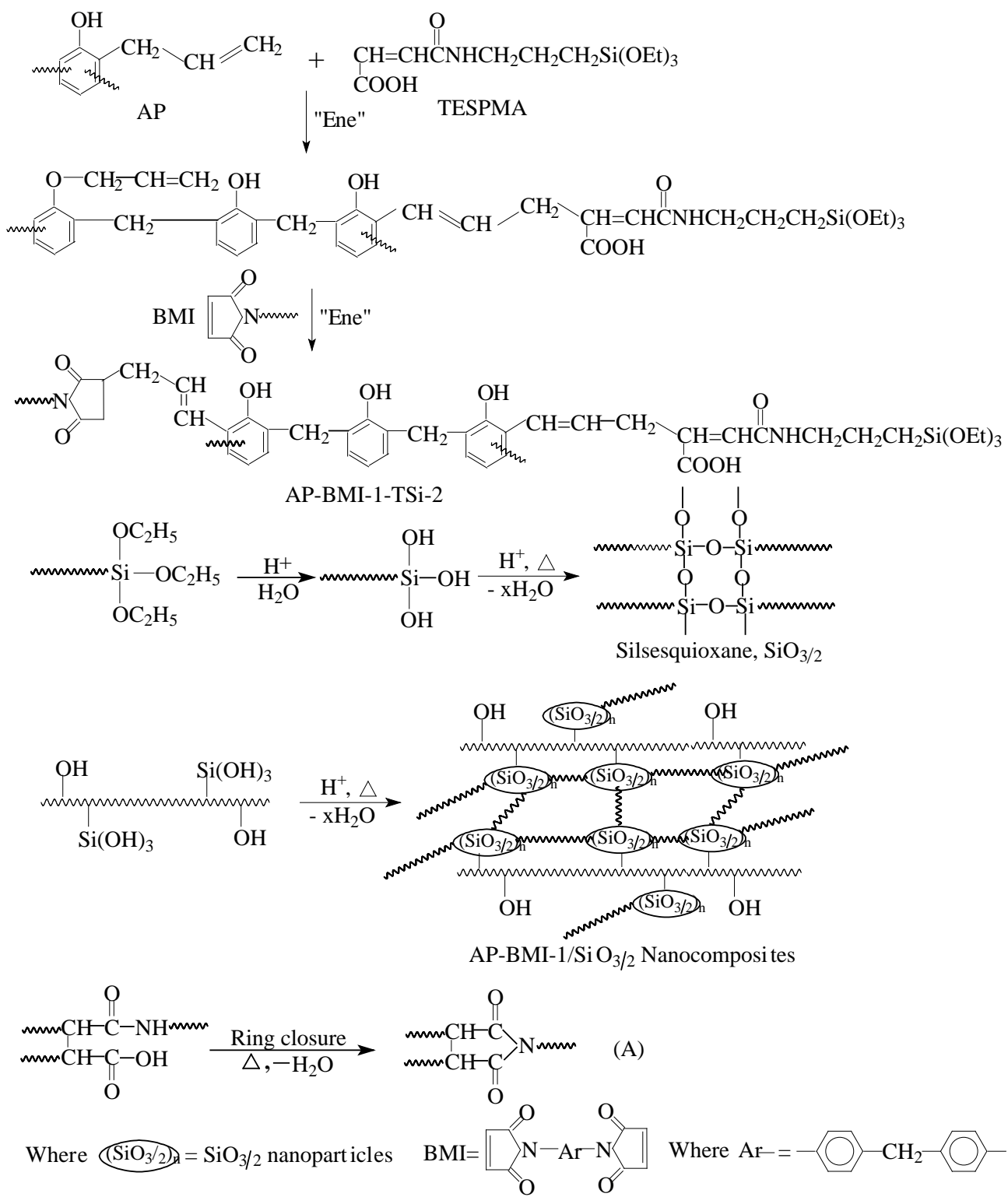

Scheme 1. The preparation of AP-BMI-1/SiO$_{3/2}$ nanocomposites.

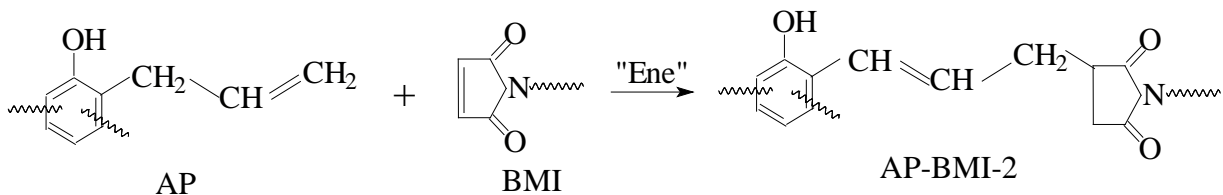

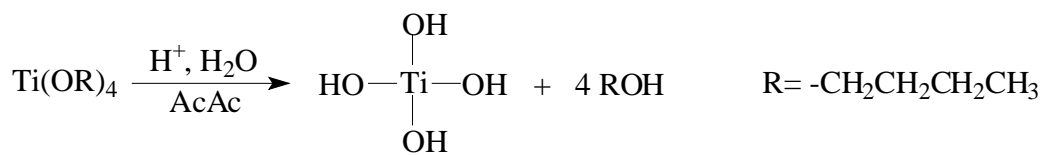

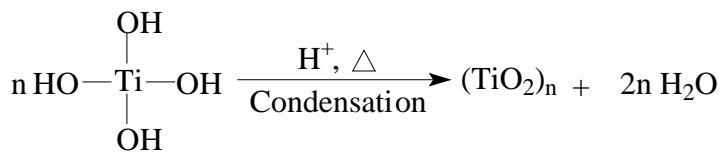

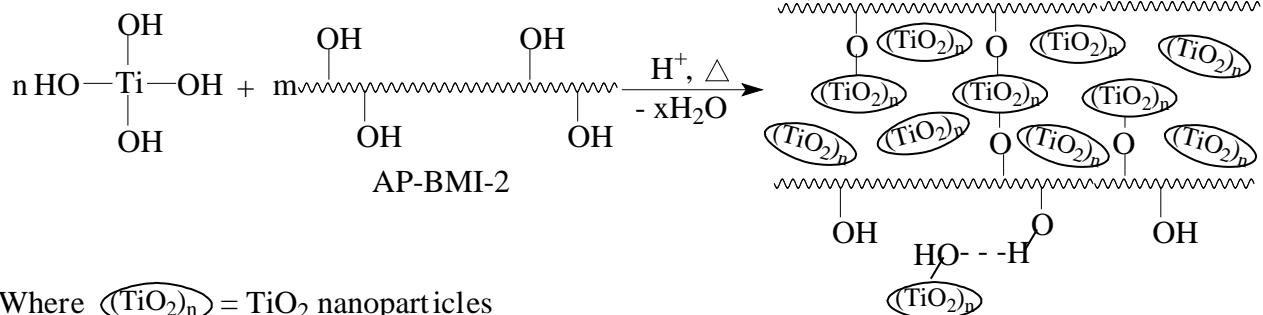

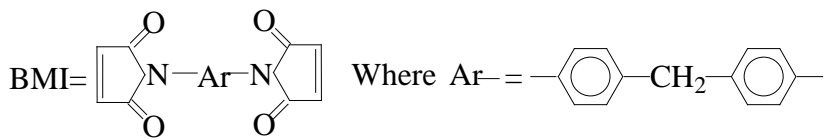

Scheme 2. The preparation of AP-BMI-2/TiO$_2$ nanocomposites.

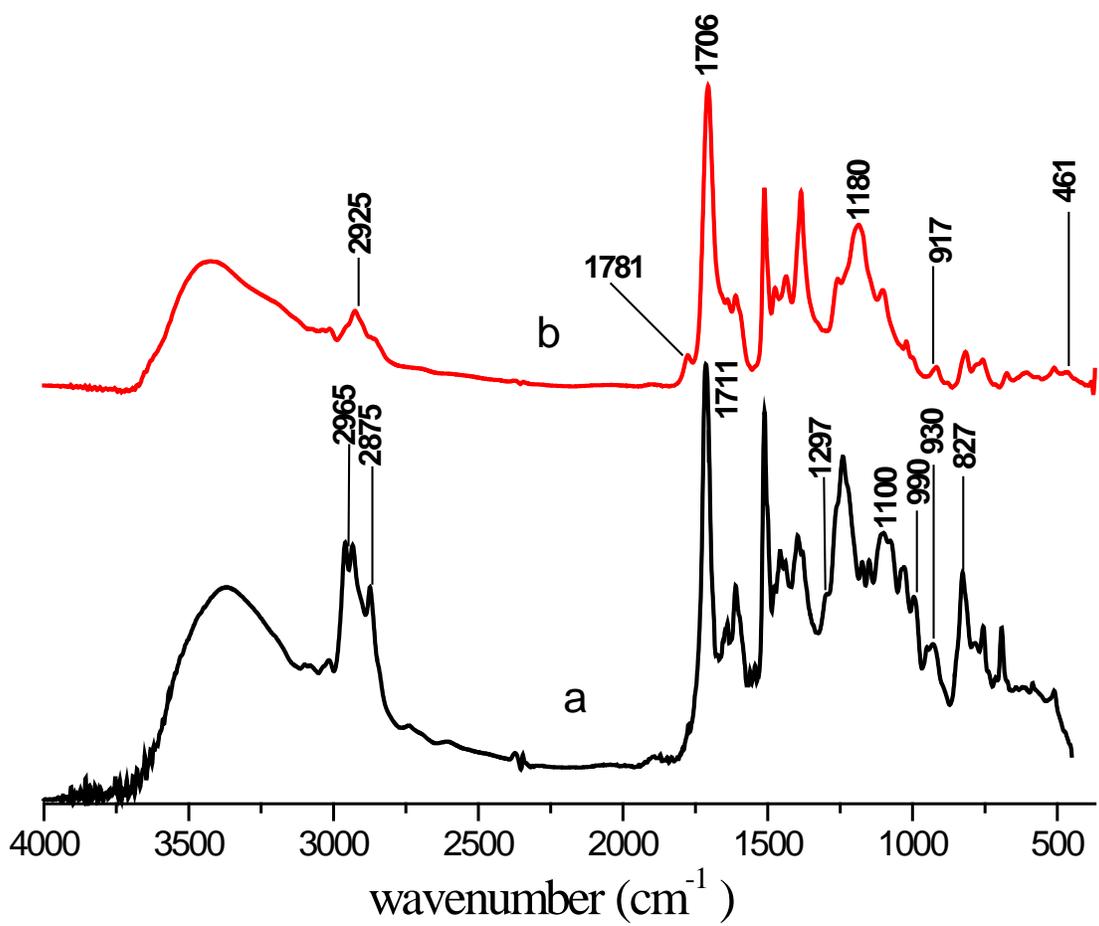

Fig 1. FT-IR spectra for the AP-BMI resin/SiO$_{3/2}$ nanocomposites: (a) AP-BMI-1-TSi; and (b) AP-BMI/Sil-2.

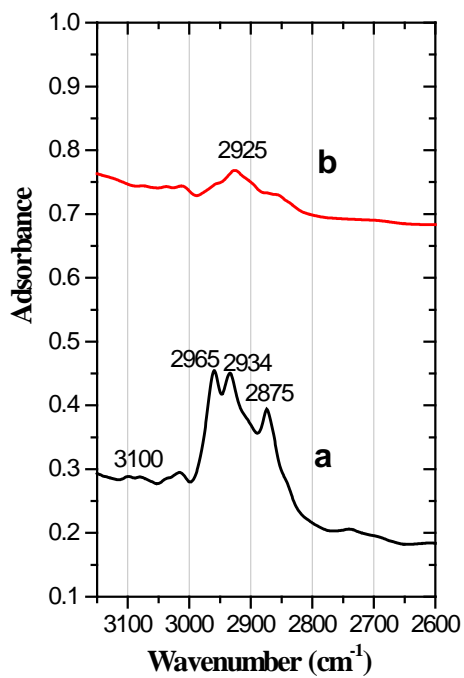
(a)

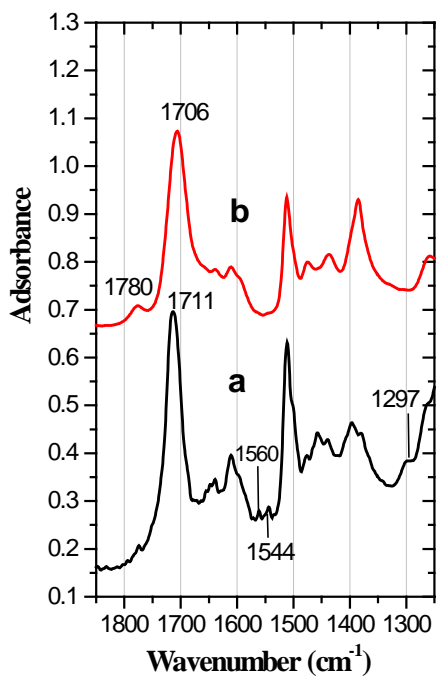
(b)

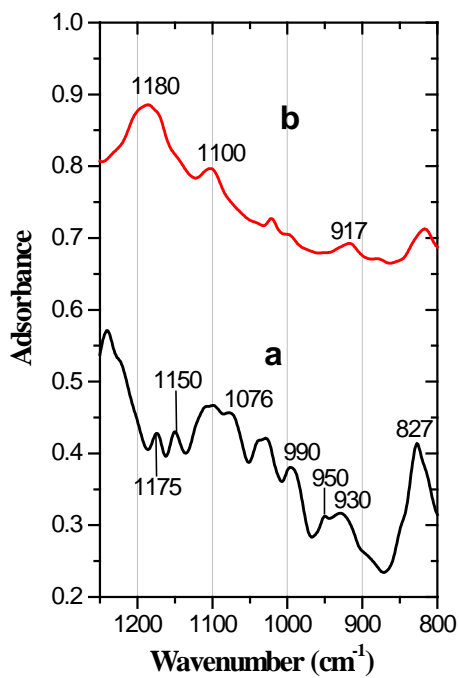
(c)

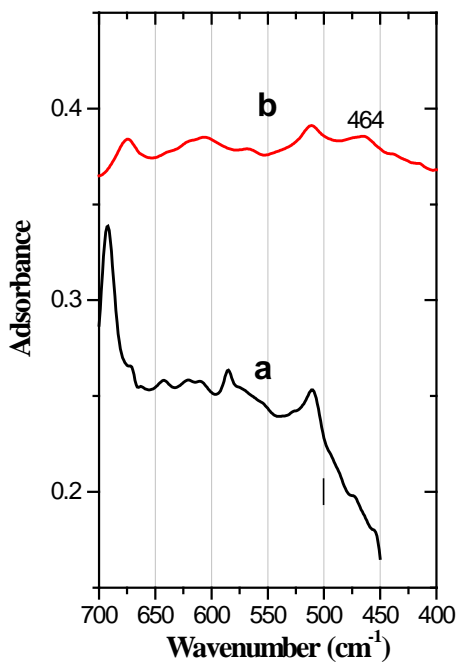
(d)

Fig 2. Magnification of Figure 1.

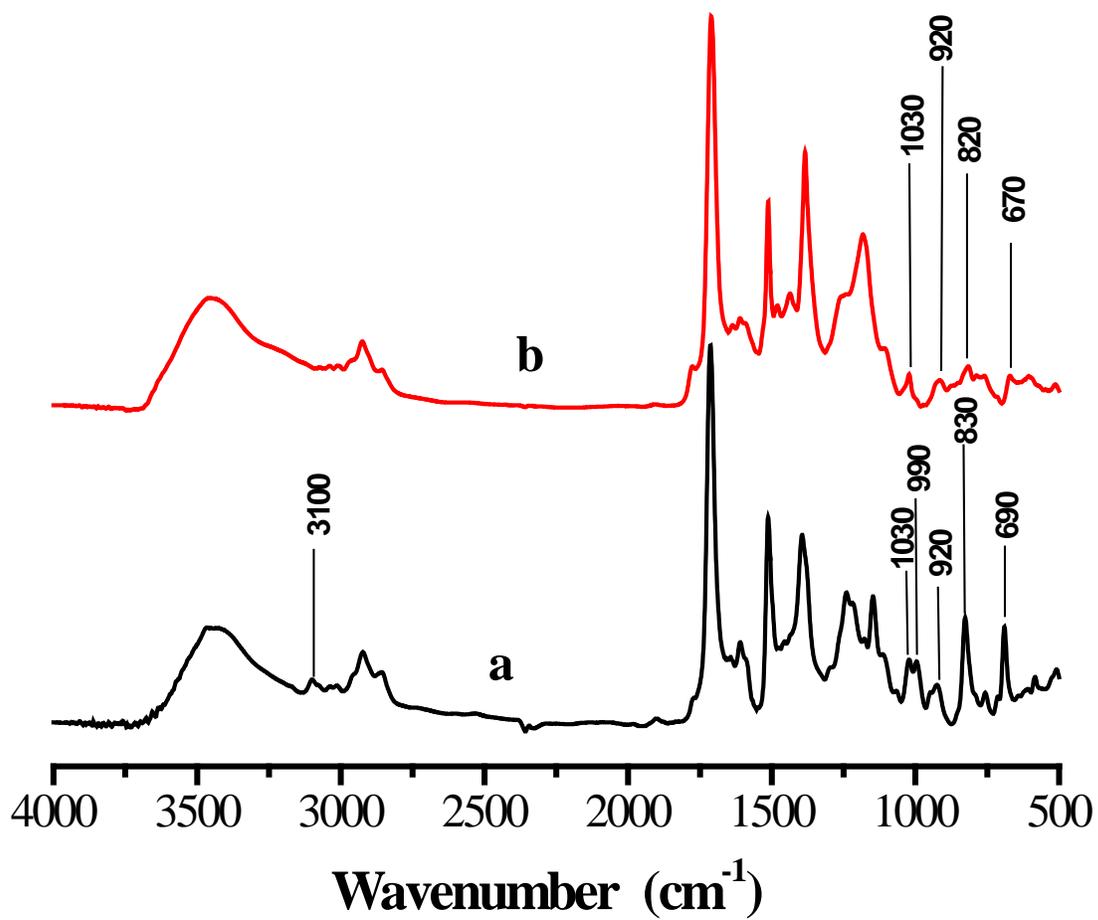

Fig 3. FT-IR spectra for the AP-BMI resin/$TiO_2$ nanocomposites: (a) AP-BMI-2; and (b) AP-BMI/Tit-2;

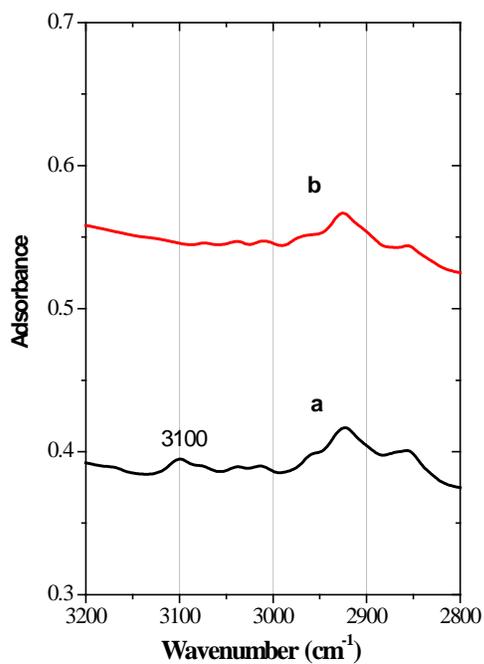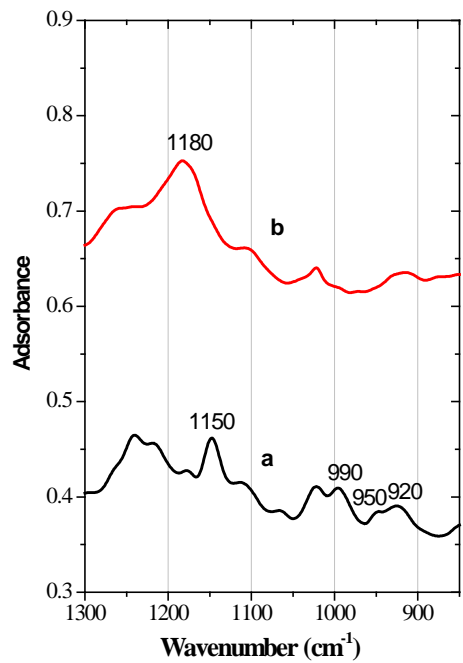

(a)

(b)

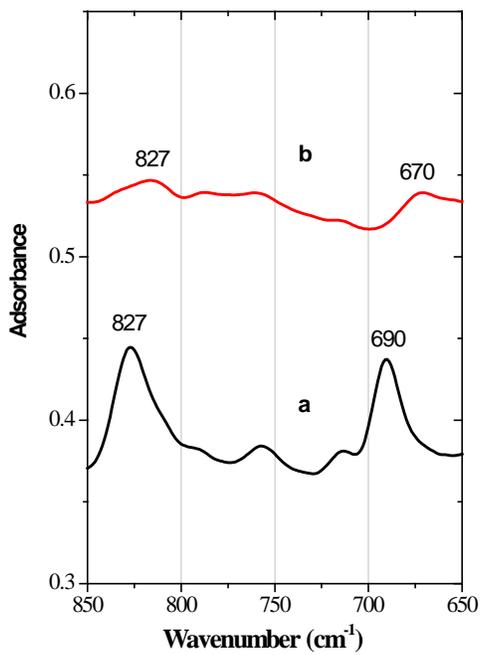

(c)

Fig 4. Magnification of Figure 3.

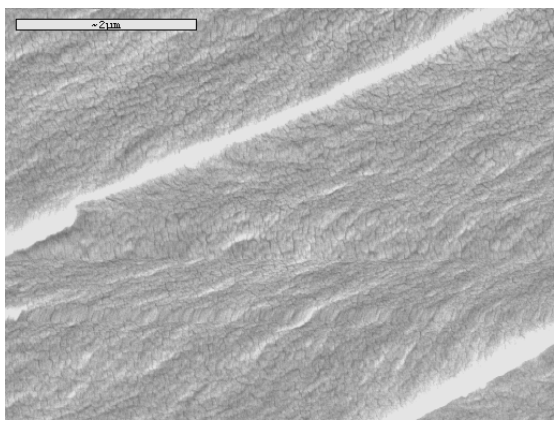
**(a)**

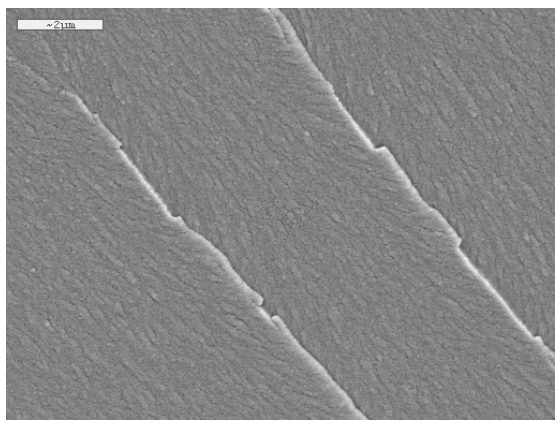
**(d)**

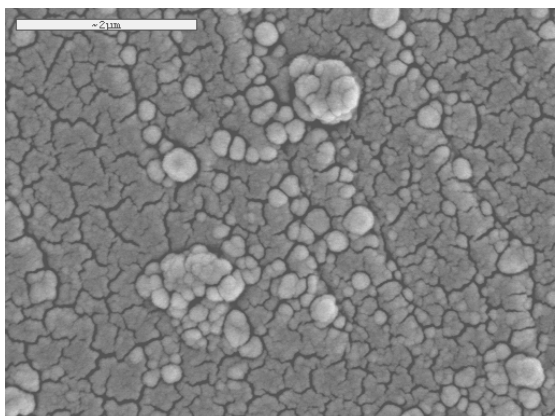
**(b)**

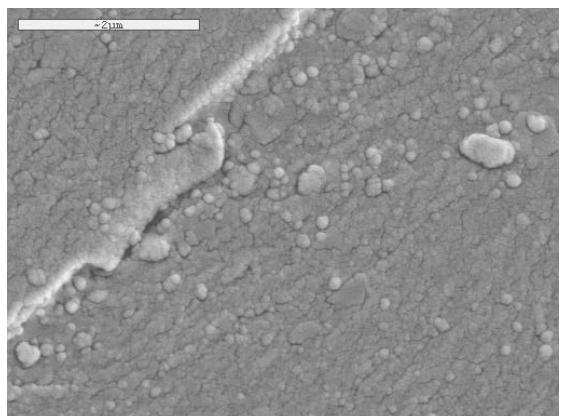
**(e)**

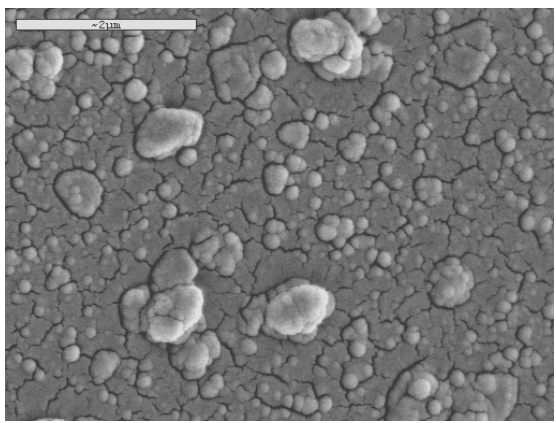
**(c)**

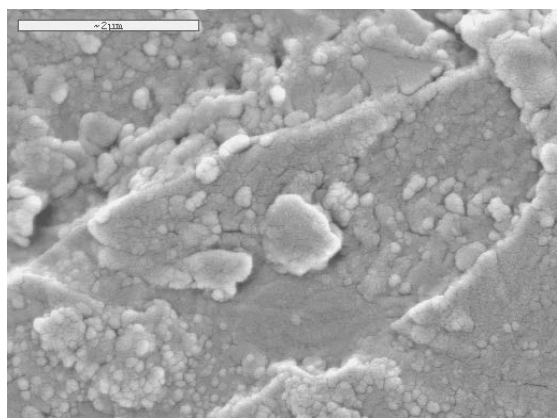
**(f)**

Fig 5. FE-SEM images for the AP-BMI resin/SiO$_{3/2}$ or TiO$_2$ nanocomposites (the scale bar is 2μm): (a) cured AP-BMI-1; (b) AP-BMI/Sil-2; (c) AP-BMI/Sil-4; (d) cured AP-BMI-2; (e) AP-BMI-2/Tit-2; and (f) AP-BMI-2/Tit-5.

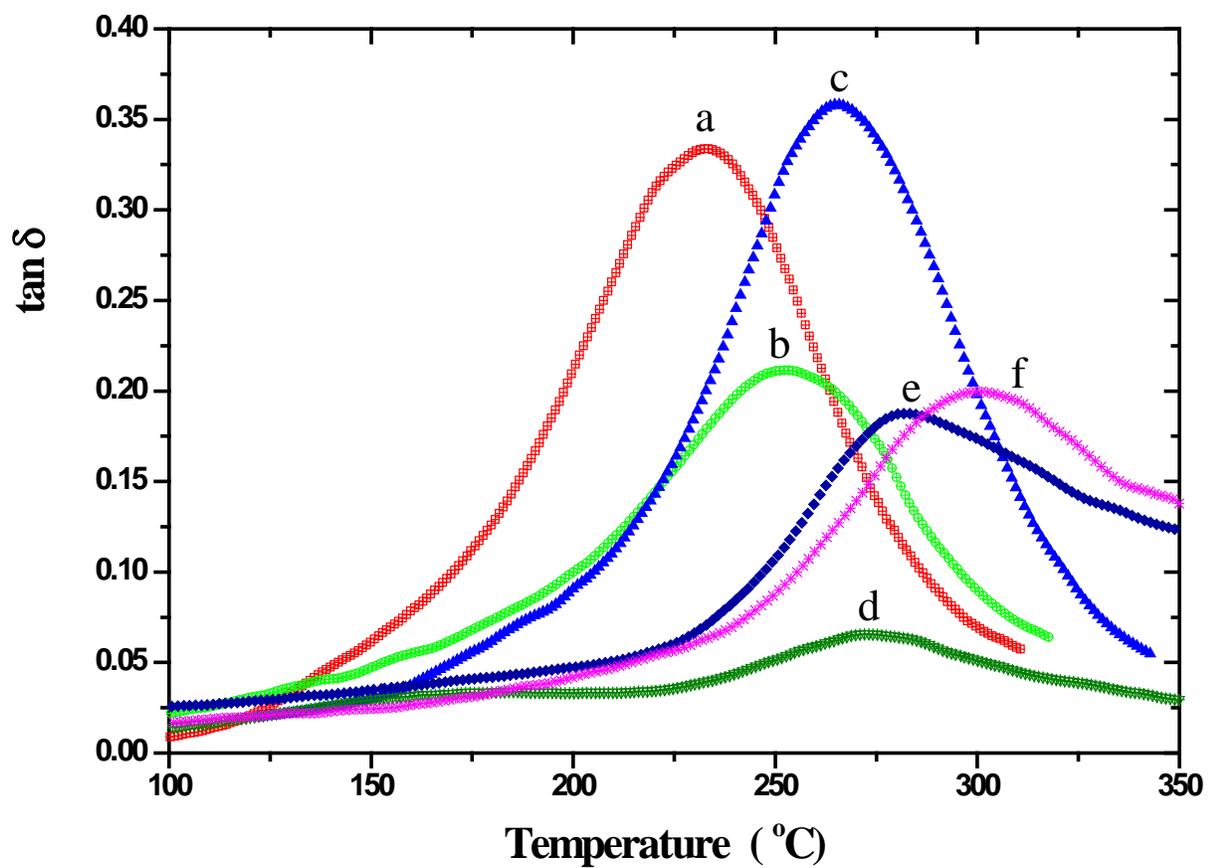

Fig 6. Glass transition temperatures of the AP-BMI resin/SiO$_{3/2}$ or TiO$_2$ nanocomposites measured by DMA. (a) cured AP-BMI-1; (b) AP-BMI/Sil-2; (c) AP-BMI/Sil-4; (d) cured AP-BMI-2; (e) AP-BMI-2/Tit-2; and (f) AP-BMI-2/Tit-5.

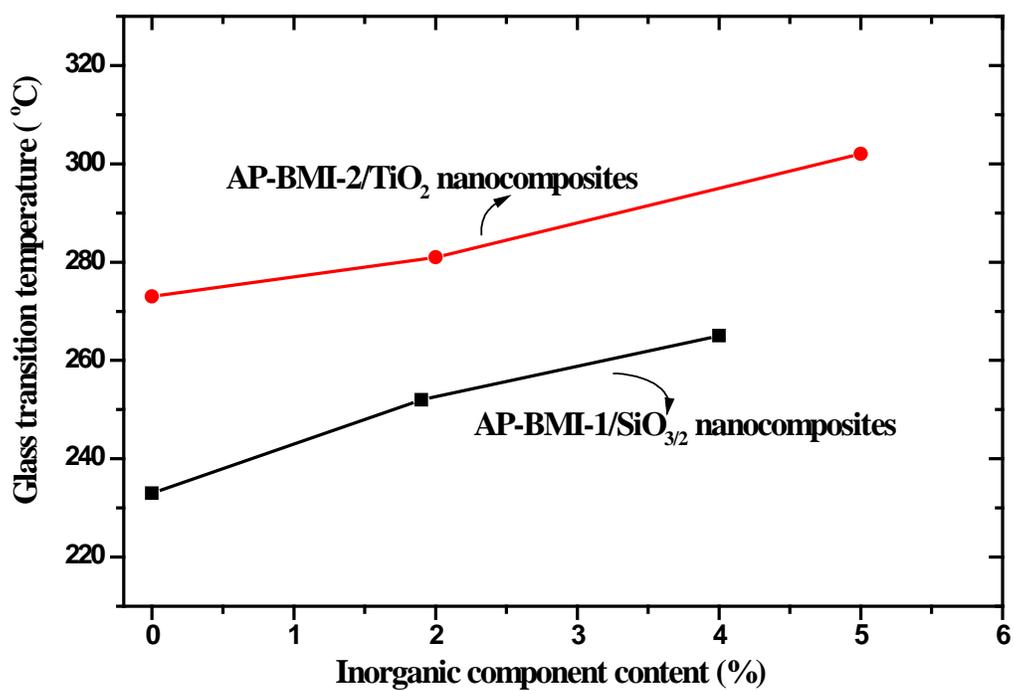

Fig 7. Glass transition temperatures of the AP-BMI resin/SiO$_{3/2}$ or TiO$_2$ nanocomposites changes with inorganic component content.

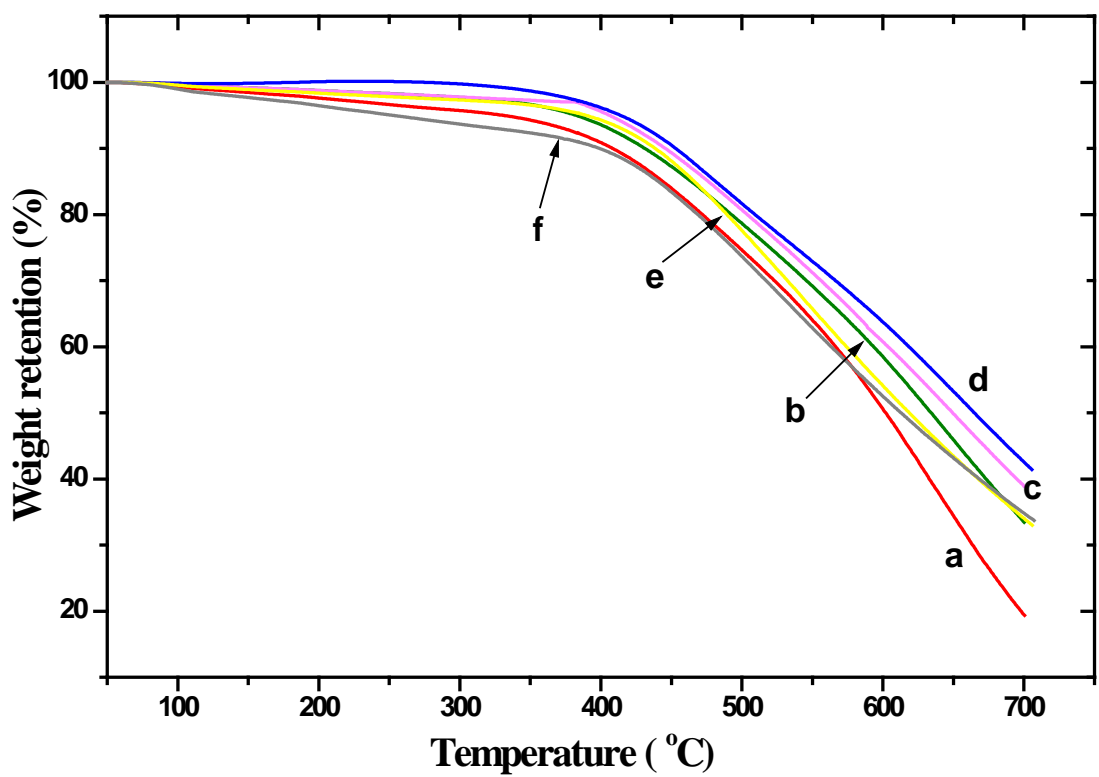

Fig 8. TGA diagrams of the AP-BMI resin/$SiO_{3/2}$ or $TiO_2$ nanocomposites: (a) cured AP-BMI-1; (b) AP-BMI/Sil-2; (c) AP-BMI/Sil-4; (d) cured AP- BMI-2; (e) AP-BMI-2/Tit-2; and (f) AP-BMI-2/Tit-5.

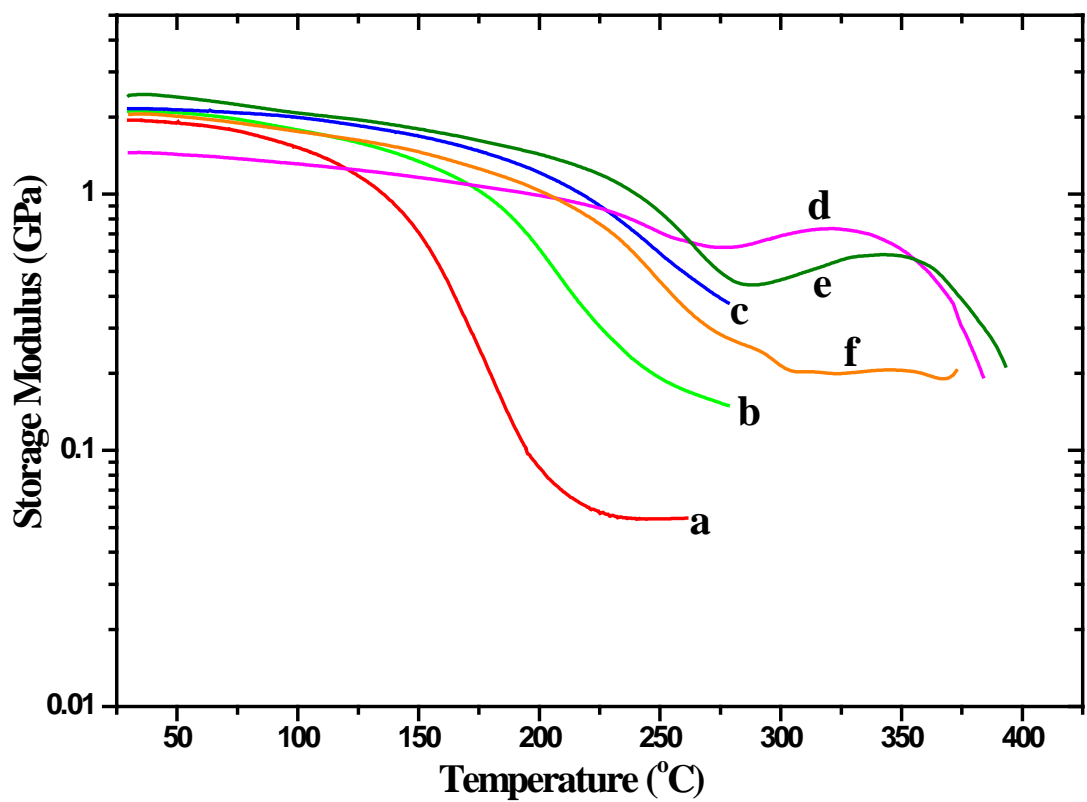

Fig 9. Change of storage modulus with temperature for the AP-BMI resin/SiO$_{3/2}$ or TiO$_2$ nanocomposites: (a) cured AP-BMI-1; (b) AP-BMI/Sil-2; (c) AP-BMI/Sil-4; (d) cured AP- BMI-2; (e) AP-BMI-2/Tit-2; and (f) AP-BMI-2/Tit-5.